\documentclass[a4paper,aps,11pt]{revtex4-1}
\usepackage{graphicx}
\usepackage{amsmath}
\usepackage{fancyhdr}
\usepackage{amssymb}
\usepackage{natbib}
\usepackage{cancel}
\usepackage{color}
\setcitestyle{round}
\begin{document}
\author{Purushottam D. Dixit}
\email{pd2447@columbia.edu}
\thanks{Corresponding author}
\affiliation{Department of Systems Biology, Columbia University, NY, NY 10032}
\title{Inferring microscopic kinetics of a Markov process using maximum caliber}
\author{Ken A. Dill}
\affiliation{Laufer Center for Quantitative Biology,\\Department of Chemistry,\\ and Department of Physics and Astronomy,\\Stony Brook University, Stony Brook, NY, 11790}
\begin{abstract}
We present a principled approach for estimating the matrix of microscopic rates among states of a Markov process, given only its stationary state population distribution and a single average global kinetic observable.  We adapt Maximum Caliber, a variational principle in which a path entropy is maximized over the distribution of all the possible trajectories, subject to basic kinetic constraints and some average dynamical observables.  We show that this approach leads, under appropriate conditions, to the continuous-time master equation and a Smoluchowski-like equation that is valid for both equilibrium and non-equilibrium stationary states. We illustrate the method by computing the solvation dynamics of water molecules from molecular dynamics trajectories.
\end{abstract}
\maketitle
\section{Introduction}
We are interested in a principled way to solve the following under-determined ``inverse'' kinetics problem.  Consider a stationary and irreducible Markov process among $i = 1,2,3, \ldots, N$ states.  Suppose you know: (a) the stationary state probability distribution, $\{ p_i \}$ of the occupancies of those states, and (b) the value $\langle w \rangle$ of some dynamical observable $w$ averaged over the ensemble of stationary state trajectories.  From these $N + 1$ quantities, we want to infer the $N \times N$ \emph{microscopic} transition rates, $\{ k_{ij} \}$ between those states.

This question is pertinent for situations such as the following.  Often molecular dynamics simulations are performed on complex systems~\citep{shaw2010atomic}, where it the stable states can be sampled more efficiently than the transitions between them, because the latter involve crossing barriers that can sometime be high.  Given the populations of the stable states, and a little experimental information about the overall rate of the process, it would be useful to estimate the full microscopic matrix of the transition rates between the states.  Other examples include how amino-acid sequences of proteins change during the evolutionary dynamics of organisms, such as the HIV virus~\citep{shekhar2013} and the collective firing patterns of neurons~\citep{schneidman2006weak}, etc. 

Here, we propose a procedure based on the principle of Maximum Caliber, a variant of the principle of Maximum Entropy, that is applicable to dynamical processes ~\citep{jaynes1980minimum,jaynes1963irreversible,Press2012}. 

First, we define \emph{the path entropy}, $\mathcal{S}$, over a given ensemble $\{ \Gamma \}$ of trajectories $\Gamma$ as:
\begin{eqnarray}
\mathcal{S} = -\sum_{\{\Gamma\}} p(\Gamma) \log p(\Gamma). \label{eq:c0}
\end{eqnarray}
Maximum Caliber is a variational principle that chooses a unique probability distribution $\{ P(\Gamma) \}$ over trajectories from all possible candidate distributions as the one that maximizes the path entropy while otherwise satisfying the relevant stationary and dynamical constraints ~\citep{jaynes1980minimum,Press2012,Stock2008,wang2005non}. 

Consider an ensemble of stationary state trajectories $\{ \Gamma \}$ of a Markov process having a total time duration $T$ where $\Gamma \equiv \cdots \rightarrow i \rightarrow j \rightarrow k \rightarrow l \rightarrow \cdots$. The Markov property implies that the probability of any particular trajectory $\Gamma$ can be expressed in terms of the transition rates $\{ k_{ij} \}$,
\begin{eqnarray}
P(\Gamma) = \cdots k_{ij} \cdot k_{jk} \cdot k_{kl} \cdots.
\end{eqnarray}
The path entropy of the above ensemble is directly proportional to the total duration $T$ of the trajectory. The path entropy per unit time $\mathcal S$ is given by~\citep{cover2012elements}
\begin{eqnarray}
\mathcal S &=& -\sum_{i,j} p_i  k_{ij}\log k_{ij}. \label{eq:c1}
\end{eqnarray}

The microscopic rates of any Markov process are subject to two types of constraints.  First, from state $i$ at time $t$, the system \emph{must land somewhere} at time $t+dt$. Second, a system in state $j$ at time $t+dt$ \emph{must arrive from somewhere}, so:
\begin{eqnarray}
\sum_j k_{ij} &=& 1~\forall~i~{\rm and}~p_j = \sum_i p_i k_{ij}~\forall~j\label{eq:st0a}
\end{eqnarray}

Third, we require one additional constraint that is \emph{global, i.e.} averaged over the entire ensemble of trajectories.  We fix the path ensemble average of some dynamical quantity $w$. The average $\langle w \rangle_{\Gamma}$ over any given stationary trajectory $\Gamma$ is given by
\begin{eqnarray}
\langle w \rangle_{\Gamma} = \frac{1}{T}\left (\dots +  w_{ij} + w_{jk} + w_{kl} + \dots \right ).\label{eq:pathensemble}
\end{eqnarray}
The path ensemble average $\langle w \rangle$ is
\begin{eqnarray}
\langle w \rangle = \sum_{\Gamma} P(\Gamma)\langle w \rangle_{\Gamma}. \label{eq:end}
\end{eqnarray}
Since $\Gamma$ is a stationary state trajectory, the path ensemble average $\langle w \rangle$  of Eq.~\ref{eq:end} simplifies to
\begin{eqnarray}
\langle w \rangle &=& \sum_{i,j} p_i k_{ij}w_{ij}. \label{eq:cn1}
\end{eqnarray}

Maximization of the path entropy subject to these three constraints can be expressed equivalently in terms of maximization of a quantity called the Caliber $\mathcal C$~\citep{Press2012}:
\begin{eqnarray}
&\mathcal C& = -\sum_{i,j} p_i  k_{ij}\log k_{ij} +  \sum_i a_i \left ( \sum_j p_i k_{ij} - p_i\right ) \nonumber \\ &+&  \sum_j l_j \left ( \sum_i p_i k_{ij} - p_j\right )- \gamma \left ( \sum_{i,j}  p_i k_{ij} w_{ij} - \langle w  \rangle \right ) \label{eq:fullC}
\end{eqnarray}
where $\gamma$ is the Lagrange multiplier associated with the constraint $\langle w \rangle$ and $\{ a_i \}$ and $\{ l_i\}$ enforce the \emph{to-somewhere} constraint and the \emph{from-somewhere} constraint, respectively.

To solve for the matrix $k_{ij}$ of rates, we take the derivative of the Caliber $\mathcal C$ with respect to $k_{ij}$ and equate it to zero.  This gives:
\begin{eqnarray}
p_i (1 + \log k_{ij}^\ast ) &=& a_i p_i + l_j p_i  - \gamma p_iw_{ij} \nonumber \\
\Rightarrow k_{ij}^\ast &=& \frac{\beta_i}{p_i} \lambda_j e^{-\gamma w_{ij}} \label{eq:pij_form}
\end{eqnarray}
where we have made the substitutions: $e^{a_i-1} = \frac{\beta_i}{p_i}$ and  $e^{l_j} = \lambda_j$.  The values $k_{ij}^\ast$ are the rates that satisfy the constraints and otherwise maximize the caliber. For simplicity of notation, we drop the superscript $^\ast$ in the remainder of this paper i.e. $k_{ij}^\ast \equiv k_{ij}$.

In this problem, the values of $p_i$ are given.  To compute the $k_{ij}$'s, we first must determine the values of the Lagrange multipliers $\beta_i$, $\lambda_j$, and $\gamma$.  We do so by substituting the constraint relations mentioned above.

\subsection{Determining the Lagrange multipliers\label{sc:deter}}
For a given value of $\gamma$, the {\it modified} Lagrange multipliers $\beta_i$ and $\lambda_j$ are determined by satisfying the {\it to-somewhere} and {\it from-somewhere} conditions indicated above. From Eqs.~\ref{eq:st0a} 
\begin{eqnarray}
&&1 = \sum_j \frac{\beta_i}{p_i} \lambda_j \mathcal W_{ij}~{\rm and~} p_j = \sum_i p_i \frac{\beta_i}{p_i} \lambda_j \mathcal W_{ij}\nonumber \\
&\Rightarrow& \frac{p_i}{\beta_i} = \sum_j \lambda_j \mathcal W_{ij}~{\rm and~} \frac{p_j}{\lambda_j} = \sum_i \beta_i \mathcal W_{ij} \label{eq:st3}
\end{eqnarray}
where $e^{-\gamma w_{ij}} = \mathcal W_{ij}$. Eq.~\ref{eq:st3} can be simplified if we define a non-linear operator $\mathcal D$ over column vectors $\bar x = \left [x_1, x_2 ,\dots \right ]^{\rm T}$ as $\mathcal D\left [\bar x\right ]_i = \frac{p_i}{x_i}$. We have
\begin{eqnarray}
\mathcal W \bar \lambda = \mathcal D \left [ \bar \beta \right ]{~\rm and~} \mathcal W^{\rm T} \bar \beta = \mathcal D \left [ \bar \lambda \right ]. \label{eq:sc}
\end{eqnarray}
where $\bar \lambda = \left [\lambda_1, \lambda_2, \dots \right ]^{\rm T}$ and $\bar \beta = \left [\beta_1, \beta_2, \dots \right ]^{\rm T}$ are the column vectors of Lagrange multipliers. 

For a particular value of the Lagrange multiplier $\gamma$, Eqs.~\ref{eq:sc} can be numerically and self-consistently solved for $\{ \beta_i \}$ and $\{ \lambda_i \}$.  In practice, we choose an appropriate $\gamma$ by first constructing transition rates $\{ k_{ij} \}$ for multiple values of $\gamma$ (see Eq.~\ref{eq:pij_form}) and chosing the value of $\gamma$ which satisfies
\begin{eqnarray}
\sum_{i,j} p_i k_{ij} w_{ij} = \sum_{i,j} \beta_i \lambda_j e^{-\gamma w_{ij}}w_{ij}  &=&\langle w\rangle
\end{eqnarray}
where $\langle w \rangle$ is the prescribed value of the ensemble average of the dynamical quantity $w$.

\section{An illustration: Computing the dynamics of a solvation shell from simulated populations.}

\begin{figure}[h]
        \includegraphics[scale=0.15]{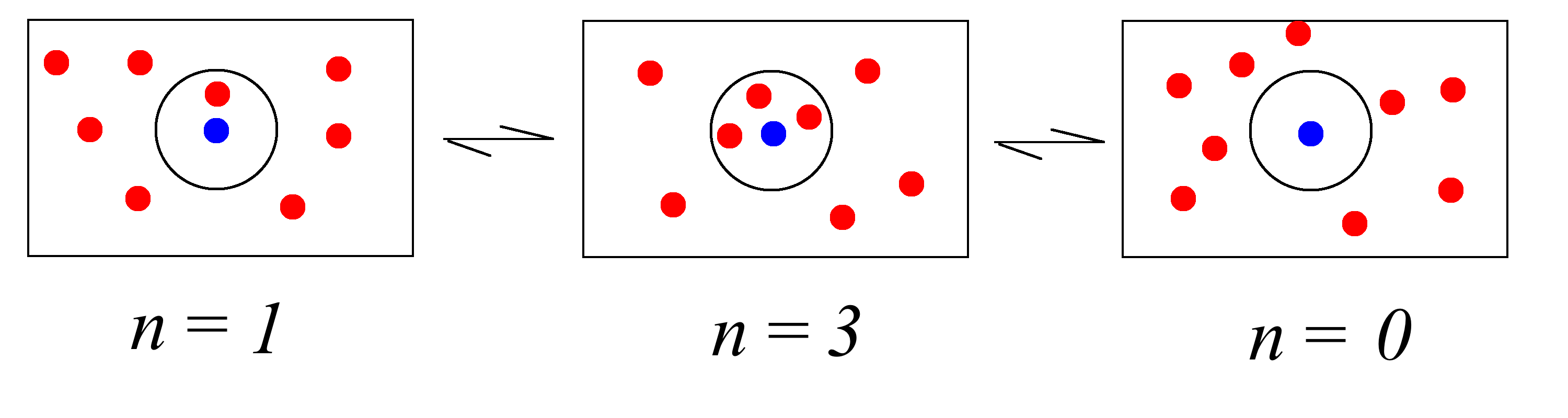}
        \caption{The hydration shell (black circle) around a central water molecule (blue disc) is dynamically populated by other water molecules in the bulk solvent medium (red discs). The probability, $p(n)$ that the hydration shell has exactly $n$ water molecules is a key quantity in determining the solvation free energy of liquid water.\label{fg:schem}}
\end{figure}

We now illustrate how the present MaxCal method can be used to take a stationary-state distribution and a global dynamical constraint and to infer microscopic kinetics.  Consider a shell of solvating water molecules surrounding a single water molecule. The number, $n(t)$, of water molecules in the hydration shell is a quantity that fluctuates with time $t$ (see Fig.~\ref{fg:schem}).  We want to compute how fast the water molecules enter or exit the solvation shell.  If the time interval $dt$ is small, $n(t)$ and $n(t+dt)$ will be statistically correlated. Here, we construct a Markov process to model the time series $\{ n(t) \}$. We will require the Markov process to reproduce a)  the stationary distribution $p(n)$ that is observed in molecular dynamics simulations, and b) the average change in occupancy $\Delta$ per time step of duration $dt$, a path ensemble average. We have
\begin{eqnarray}
\Delta &=& \langle |n(t+dt) - n(t)| \rangle = \sum_{i,j} |i-j| \cdot p(i)k_{ij} \label{eq:dcon}
\end{eqnarray}
where $n(t) = i$ and $n(t+dt) = j$.

\begin{figure}
	\begin{center}
         \includegraphics[scale=1]{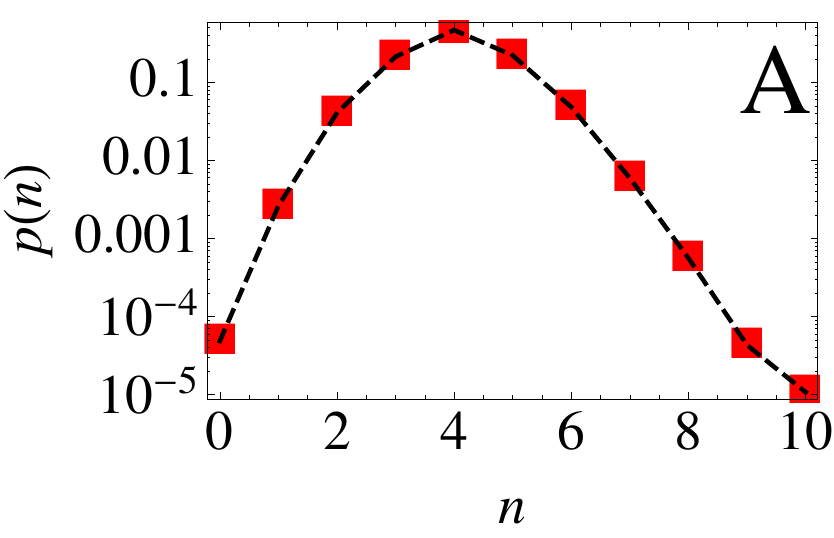}
	\includegraphics[trim = 0mm 10mm 0mm 0mm, scale=1]{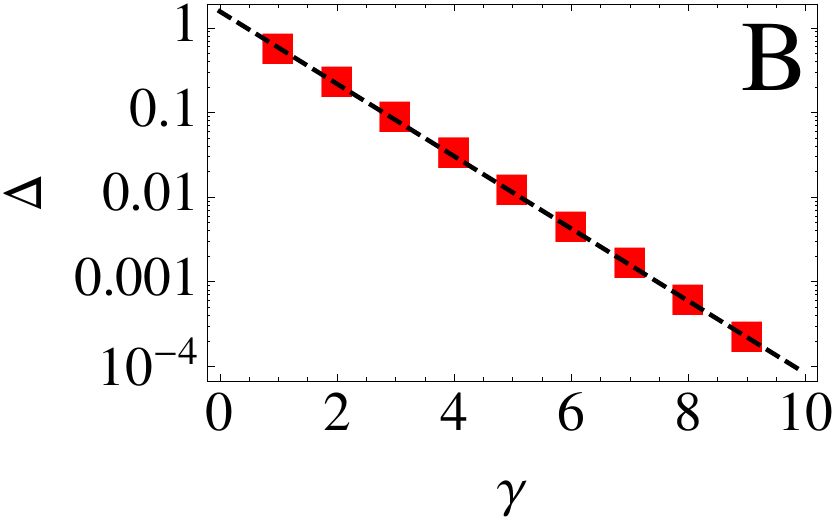}	
	\end{center}	
        \caption{Panel A: The stationary distribution $p(n)$ of the number of water molecules in the hydration shell of radius $r=3.2$~\AA~of a reference water molecule.  Panel B: The dependence of the ensemble average of change in water occupancy number $\Delta =\langle |n(t+dt) - n(t)| \rangle$ on the Lagrange multiplier $\gamma$. We see that  $\Delta$ depends exponentially on $\gamma$. A higher $\gamma$ implies slower dynamics and vice versa. We choose $\gamma = 3.29$ to match the observed $\Delta \approx 0.0629$ in the molecular dynamics simulation.\label{fg:stat}}
\end{figure}

From a trajectory sampled at every 5 fs from an MD simulation of liquid water (see appendix {\bf III} for details of the simulation),  we estimate the roughly normal stationary distribution $p(n)$ (see in panel A of Fig.~\ref{fg:stat})~\citep{asthagiri:cpl,merchant2009thermodynamically}. When we constrain the observed stationary distribution $p(n)$ and the mean jump size $\Delta$, the transition rate $k_{ij}$ for a transition $n(t) = i \rightarrow  n(t+dt) = j$ is given by (see Eq.~\ref{eq:pij_form}),
\begin{eqnarray}
        k_{ij} &=& \frac{\beta_i \lambda_je^{-\gamma |i-j|}}{p(i)}. \label{eq:model}
\end{eqnarray}

For a given value of $\gamma$, we determine the Lagrange multipliers $\beta_i$ and $\lambda_i$ from Eqs.~\ref{eq:sc} above.  In order to determine the Lagrange multiplier $\gamma$ which dictates the rate of transition between states, we first construct Markov processes for different values of $\gamma$. Panel B of Fig.~\ref{fg:stat} shows that the path ensemble average of the change in occupation number per unit time step $\Delta$ is exponentially decreasing with $\gamma$. From trajectories sampled at every 5 fs from the MD simulation, we find that experimental trajectory average $\Delta_{\rm expt} = \overline{|n(t+dt) - n(t)|} \approx 0.0629$ which corresponds to $\gamma \approx 3.29$. From here onwards, we use $\gamma = 3.29$ and construct the transition rates $\{k_{ij} \}$ (see Eq.~\ref{eq:model}). Note that the path ensemble average $\Delta$ and consequently the Lagrange multiplier $\gamma$, depend on the time interval $dt$ between two observation ($dt = 5$ fs here).  
%
\begin{figure*}            
		\includegraphics[scale=0.98]{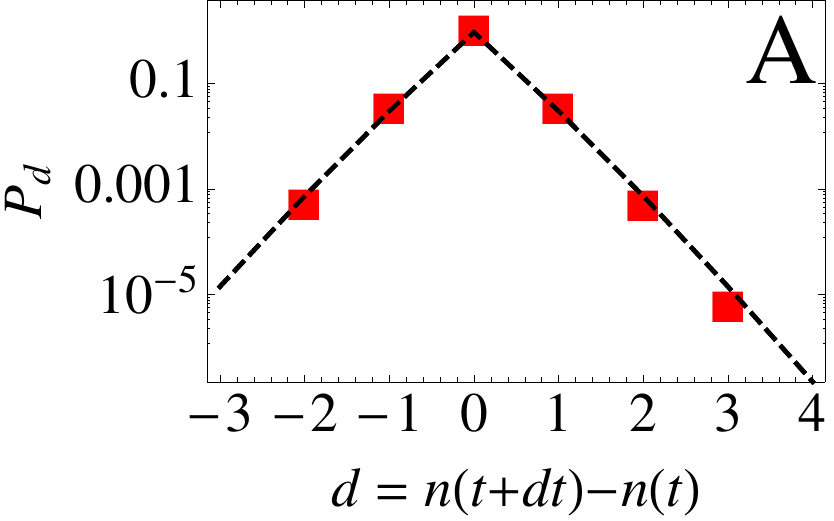}            
		\includegraphics[scale=0.915]{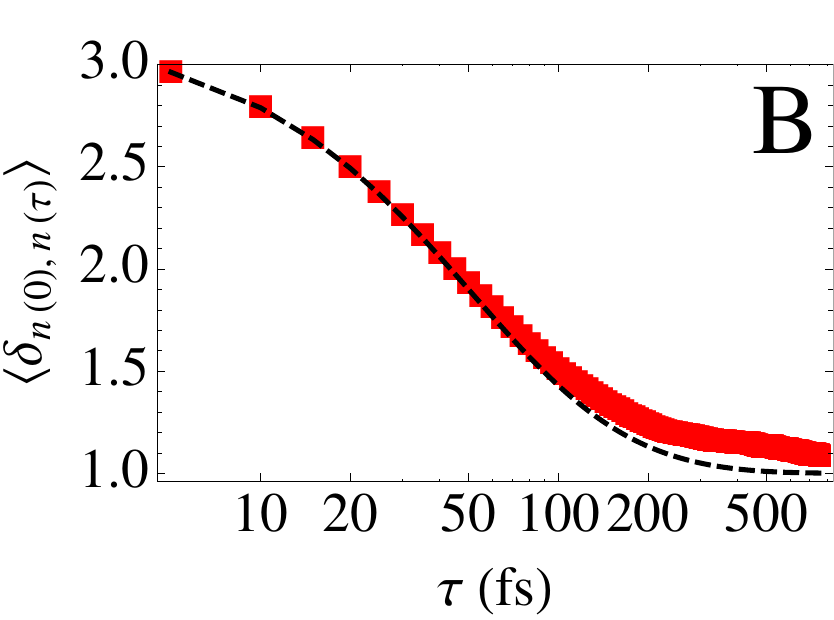}
        		\includegraphics[scale=1]{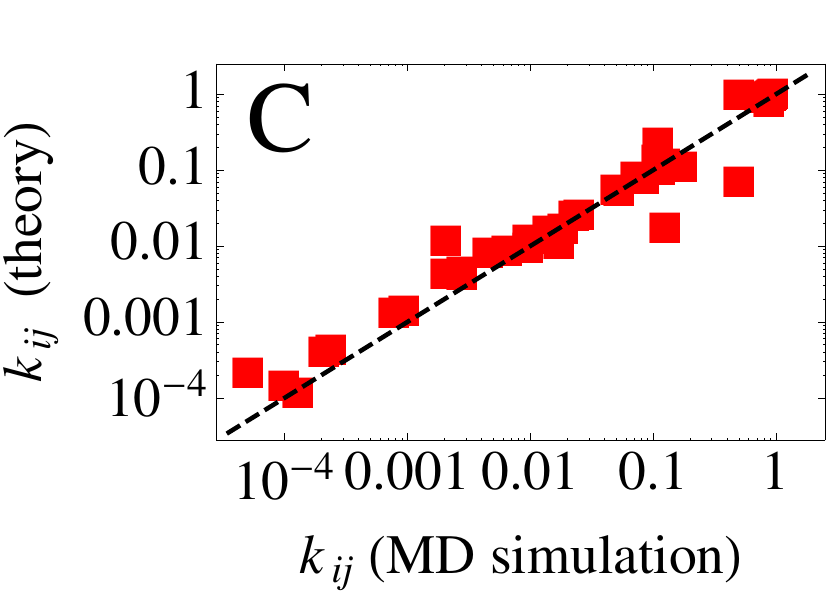}
        \caption{Panel A: The probability $P_d$ of jump size estimated from the trajectory derived from MD simulation (red squares) is compared to the one predicted using the transition rates of the Markov process (dashed black line).   Panel B: The normalized occupancy autocorrelation $\langle \delta_{n(0),n(\tau)} \rangle$ as estimated from the MD trajectory and as predicted from the transition rates of the Markov process. Panel C: We directly compare the transition rates $k_{ij}$ for the probability of transition $i\rightarrow j$ empirically obtained from the MD trajectory to the ones predicted by using Eq.~\ref{eq:model}. \label{fg:results}}
\end{figure*}

From the Markov process constructed with $\gamma = 3.29$ (see above), we now compute various dynamical quantities: a) the probability $P_d$ of jump size $d$, b) the occupancy autocorrelation $\langle \delta_{n(0),n(\tau)} \rangle$, and c) the transition probabilities $k_{ij}$, and we compare to those obtained directly from the MD simulation trajectory.  In general, the MaxCal method will be of value when rates are hard to simulate, such as for large kinetic barriers.  Here, we are just illustrating with a toy problem for which we can determine the rates independently from the simulations.

From the long simulation trajectory, the probability $P_d$ of jump size is estimated as the histogram of $d = n(t+dt)-n(t)$. Here $d$ could be both positive and negative. $P_d$ is given by
\begin{eqnarray}P_d &=& \sum_n p(n) k_{n,n+d}.
\end{eqnarray}
The normalized occupancy autocorrelation is simply the joint probability that $n(t)$ and $n(t+\tau)$ are equal. It is given by
\begin{eqnarray}
\langle \delta_{n(0),n(\tau)} \rangle &=&\frac{\sum_n p(n) K({\tau})_{nn}}{\sum_n p(n)^2} 
\end{eqnarray}
where $K({\tau}) = k^{\tau}$ is the $\tau^{\rm th}$ power of the matrix of transition rates $\{ k_{ij} \}$.

In Fig.~\ref{fg:results} we plot $P_d$, $\langle \delta_{n(0),n(\tau)} \rangle$, and $\{ k_{ij} \}$ estimated from the molecular dynamics trajectory and compare them to our predictions from Markov modeling. Even though we constrained only the mean value $\Delta = \langle |d| \rangle$ of $P_d$, the modeled Markov process captures the entire distribution $P_d$ with  high accuracy. Similarly the occupancy correlation $\langle \delta_{n(0),n(\tau)} \rangle$ is also reproduced with  high accuracy even though we did not utilize {\it any} information about it when infering the transition rates of the Markov process. Moreover, our modeling also accurately captures the individual transition rates $\{ k_{ij} \}$ over 4 orders of magnitude.

\section{Continuous time and continuous space limits}
Above, we have described discrete Markov processes.  But, it is readily shown that the method can also be applied to continuous processes.  Consider a Markov system whose states are points on a multi-dimensional discrete lattice with spacing $dx$ in each dimension. Let $d_{ij}$ be the manhattan distance between two states $i$ and $j$. A wide variety of systems belong to this class including discretized brownian walks, lattice polymers, and spin glasses. 

Assume that the system evolves continuously in time $t$ but that we observe it only at a regular time interval $dt \ll \tau$, where $\tau $ is the characteristic time constant of the system. Let us construct a Markov process by constraining the path ensemble average $\langle d_{ij} \rangle$, the average displacement per time step.
%

In appendix {\bf I} we show that as $dt \rightarrow 0$, the Markov process described above is expressed by the master equation
\begin{eqnarray}
\frac{d\bar q(t)}{dt}= -K^{\rm T} \bar q(t). \label{eq:maseq0}
\end{eqnarray}
The transition rates $i \neq j$ are given by
\begin{eqnarray} K_{ij} &=& \mu \sqrt{\frac{p_j}{p_i}}~{\it if}~d_{ij} = dx~{\rm and~} 0~{\it if}d_{ij} > dx \label{eq:maseq1}\\
 K_{ii} &=& -\sum_{j\neq i}  K_{ij}. \label{eq:maseq2}
\end{eqnarray}
Here $\bar q(t) = [q_1(t) , q_2(t), \dots ]$ denotes the instantaneous probability distribution. Intriguingly, the transition rates $K_{ij}$ are functions of probability amplitudes $\sqrt{p_j}$ and $\sqrt{p_i}$. We constrast this observation with the well known Glauber dynamics~\citep{glauber1963time} for Ising-like systems where transtion rates are functions of probabilities themselves. Given that the continuous-space limit of the master equation Eq.~\ref{eq:maseq0} is the Smoluchowski equation (see below),  we identify Eq.~\ref{eq:maseq0} as the unique form of the discrete-space Smoluchowski equation.

The continuous-time continuous space limit of Eq.~\ref{eq:maseq1} as $dx\rightarrow 0$ is the so-called Smoluchowski equation for interacting degrees of freedom. In appendix {\bf II} we show that Eq.~\ref{eq:maseq1} reduces to
\begin{eqnarray}
\frac{\partial q(\bar X;t)}{\partial t} = D \nabla^2 q(\bar X;t) - D \nabla \cdot \left [q(\bar X; t)\cdot  \overline{\mathcal F}(\bar X)  \right].\label{eq:smol}
\end{eqnarray}
Here, $q(\bar X;t)$ is the instantaneous probability density of the state space point $\bar X$, $\phi(\bar X) = - \log p_{ss}(\bar X)$ is the statistical field that corresponds to the stationary state density $p_{ss}(\bar X)$,  and $D$ is an effective diffusion constant that sets the time scale for $q(\bar X; t)$ to reach the steady state density $p_{\rm ss}(\bar X)$.  We have denoted $\overline{ \mathcal F}(\bar X) = -\nabla \phi(\bar X)$ as a restoring {\it force}.   Note that Eq.~\ref{eq:smol} is valid even if $\phi(\bar X)$ is a dissipative field corresponding to a non-equilibrium steady state.

%


\section{Discussion and Summary}

We have presented here a variational approach that computes $N \times N$ microscopic rate coefficients of a Markov process, given only knowledge of a stationary state population distribution and one trajectory-averaged dynamical property.  In this approach, we maximize a path entropy subject to constraints.  We show that this method correctly gives dynamical quantities on an example of molecular dynamics simulations of a water solvation shell around a single water molecule.  This method may be useful for analyzing dynamical data from MD simulations~\citep{shaw2010atomic}, single-molecule experiments such as on ion channels~\citep{hille2001ion}, dynamics of neuron firing~\citep{schneidman2006weak}, and the dynamics of protein-sequence evolution~\citep{shekhar2013}, for example.

\begin{acknowledgments}
KD would like to thank YYY. PD thanks Mr. Manas Rachh and Mr. Karthik Shekhar for numerous discussions about the topic and a critical reading of the manuscript.
\end{acknowledgments}

\newpage

\section*{Appendix I: Continuous time limit}
Since Eq.~\ref{eq:fullC} is convex in transition rates $k_{ij}$, there exists a unique matrix of transition rates $\{ k_{ij} \}$ that maximizes the Caliber among all transition rate matrices $\{ k_{ij} \}$ that satisfy the imposed stationary and dynamical constraints. In the main text, we showed that the transition rates are given by
\begin{eqnarray}
k_{ij} = \frac{\beta_i}{p_i} \lambda_j e^{-\gamma w_{ij}}
\end{eqnarray}
where for a given value of $\gamma$,  the Lagrange multipliers $\{ \beta_i \}$ and $\{ \lambda_j\}$ are determined by satisfying the normalization and and stationary constraints of Eq.~\ref{eq:sc}.

If the constrained quantity $w_{ij}$ is symmetric in $i$ and $j$ for a transition $i \rightarrow j$ i.e. $w_{ij} = w_{ji}~\forall~i$ and $j$, we have  
\begin{eqnarray}
\mathcal W_{ij} = e^{-\gamma w_{ij}} = \mathcal W_{ji} = e^{-\gamma w_{ji}} \Rightarrow \mathcal W = \mathcal W^{\rm T}
\end{eqnarray}
From Eq.~\ref{eq:sc} it follows that
\begin{eqnarray}
\mathcal W \bar \lambda = \mathcal D[\bar \beta]{~\rm and~} \mathcal W \bar \beta = \mathcal D[\bar \lambda].
\end{eqnarray}
Identifying $\mathcal W^{-1} \mathcal D = \mathcal G$ as a non-linear operator on column vectors, we have
\begin{eqnarray}
\bar \lambda &=& \mathcal G[\bar \beta]{~\rm and~}\bar \beta = \mathcal G[\bar \lambda]. \label{eq:scs}
\end{eqnarray}
Notice that if $\{ \bar \lambda,  \bar \beta \}$ is a solution of Eq.~\ref{eq:scs}, so is $\{\bar \beta, \bar \lambda \}$. In other words, {\it since} $\{ k_{ij} \}$ are uniquely determined from the maximization of Eq.~\ref{eq:fullC}, we must have
\begin{eqnarray}
\lambda_i \beta_j &=& \beta_i \lambda_j~ \forall~i~{\rm and}~j. \label{eq:symm}
\end{eqnarray}
From Eq.~\ref{eq:symm} it follows that $\lambda_i = \eta \beta_i$ for some non-zero constant $\eta$. We have
\begin{eqnarray}
\bar \lambda = \mathcal W^{-1} \mathcal D[\bar \beta] \Rightarrow \mathcal G[\bar \beta] = \eta \bar \beta.
\end{eqnarray}
In other words, $\bar \beta$ is the eigenvector of the non-linear operator $\mathcal G =\mathcal W^{-1} \mathcal D$ with eigenvalue $\eta$ and $\bar \lambda = \eta \bar \beta$. Now it follows that
\begin{eqnarray}
k_{ij} &=& \frac{\eta \lambda_i \lambda_j e^{-\gamma w_{ij}}}{p_i} \label{eq:sym}
\end{eqnarray}

\vspace{3mm}
Now consider a  discrete-time discrete-state Markov process over states $\{i \}$. Assume that the states are points on a multi-dimensional lattice with spacing $dx$.  Let $d_{ij}$ be the manhattan distance between states $i$ and $j$.  Let us constrain the path ensemble average of the movement $\langle d_{ij} \rangle$ in a single step over long trajectories. Since $d_{ij}$ is symmetric, from Eq.~\ref{eq:sym} we have
\begin{eqnarray}
k_{ij} &=& \frac{\eta \lambda_i \lambda_je^{-\gamma d_{ij}}}{p_i}. \label{eq:metric1}
\end{eqnarray}
The Lagrange multipliers $\bar \lambda$ are determined by solving Eq.~\ref{eq:sc}
\begin{eqnarray}
\mathcal D[\bar \lambda] &=& \eta \mathcal W \bar \lambda.
\end{eqnarray}
Applying $\mathcal D$ on both sides and recognizing that $\mathcal D[c\bar x] = \mathcal D[\bar x]/c$ for any non-zero $c$,
\begin{eqnarray}
\eta \bar \lambda &=& \mathcal D [ \mathcal W\bar \lambda].
\end{eqnarray}

Now consider the case where $\gamma$ is positive and large we write $\epsilon = e^{-\gamma dx} \ll 1$ where $dx$ is the minimum pairwise separation between all states $i$ and $j$. Here, $\epsilon$ is small. 

When $\gamma$ is positive and large, the Markov process realistically only visits nearest neighbor states. For example, for any state $i$, the realistically accessible nearest neighbor states are all states $j$ such that $d_{ij} = dx$.

Since $d_{ii} = 0$, we can write $\mathcal W \approx \mathcal I + \epsilon \Delta$ where $\mathcal I$ is the identity matrix. $\Delta$ is a matrix of connectivity of nearest neighbors, $\Delta_{ij} = 1$ {\it iff} $d_{ij} = dx$ and zero otherwise. We have
\begin{eqnarray}
\eta \bar \lambda &=& \mathcal D [\bar \lambda + \epsilon \bar f] \label{eq:dexpand}
\end{eqnarray}
where
\begin{eqnarray}
\bar f &=& \Delta \bar \lambda.
\end{eqnarray}
Expanding the right hand side of Eq.~\ref{eq:dexpand} and keeping terms of order up to $\epsilon$,
\begin{eqnarray}
\eta \lambda_i &\approx& \frac{p_i}{\lambda_i} - \frac{\epsilon p_i f_i}{\lambda_i^2}. \label{eq:cubic}
\end{eqnarray}
Notice that since $\Delta_{ii} = 0$, $f_i$ does not directly depend on $\lambda_i$. We solve Eq.~\ref{eq:cubic} for $\lambda_i$ and neglect terms of order higher than $\epsilon$. We have
\begin{eqnarray}
\lambda_i &\approx& \sqrt{\frac{p_i}{\eta}} - \frac{\epsilon f_i}{2} \\
\Rightarrow \lambda_i &=& \frac{1}{\sqrt{\eta}} \left ( \sqrt{p_i} - \frac{\epsilon}{2}\sum_j \Delta_{ij} \sqrt{p_j} \right ) \label{eq:final1}
\end{eqnarray}
Recall that
\begin{eqnarray}
k_{ij} &=& \frac{\eta \lambda_i \lambda_j e^{-\gamma d(i,j)}}{p_i}  \\
&=& \frac{\eta\epsilon \lambda_i \lambda_j}{p_i}\nonumber \\~{\it iff}~d_{ij} &=& dx{~\rm and}~i\neq j{\rm ~and~0~otherwise}\label{eq:final2}
\end{eqnarray}
Substituting $\lambda_i$ from Eq.~\ref{eq:final1} into Eq.~\ref{eq:final2} and retaining terms up to order 1 in $\epsilon$, we get
\begin{eqnarray}
k_{ij} &=& \epsilon \sqrt{\frac{p_j}{p_i}}\nonumber \\~{\it iff}~d_{ij} &=& dx {~\rm and}~i\neq j{\rm ~and~0~otherwise}\\
k_{ii} &=& 1-\sum_{j\neq i} k_{ij} \label{eq:smalls}
\end{eqnarray}

Let $\bar q(t)$ be the instantaneous probability distribution of the Markov process. From Eq.~\ref{eq:smalls}, we have
\begin{eqnarray}
q_j(t+dt) &=&\sum_i k_{ij}q_i(t).
\end{eqnarray}
Writing $\epsilon = \mu dt$ and taking the limit $dt \rightarrow 0$, we derive Eq.~\ref{eq:maseq0}.

\section*{Appendix II: Deriving the Smoluchowski equation}
Here, we will derive the Smoluchowski equation for a particle diffusing in a one dimensional landscape. The generalization Eq.~\ref{eq:smol} presented in the main text is trivially obtained from the one-dimensional equation. Let us consider a particle moving between $x = -L$ and $x=L$ in discrete steps of size $dx$. Let the stationary distribution $p_{\rm ss}(x)$ be governed by a potential $\phi(x)$ such that $p_{\rm ss}(x) = e^{-\phi(x)}$. Let us assume that the particle diffuses in continuum time on this discrete landscape. Denote the instantaneous probability density by $q(x;t)$. From Eq.~\ref{eq:maseq1}, we have
\begin{widetext}
\begin{eqnarray}
\frac{dq(x;t)}{dt} &=& \underbrace{\mu \left ( q(x+dx;t)\sqrt{\frac{e^{-\phi(x)}}{e^{-\phi(x+dx)}}} + q(x-dx;t)\sqrt{\frac{e^{-\phi(x)}}{e^{-\phi(x-dx)}}}\right )}_{\rm entry~into~state~{\it x}~from~its~neighbors} \nonumber \\&-& \overbrace{\mu \left ( q(x;t)\sqrt{\frac{e^{-\phi(x+dx)}}{e^{-\phi(x)}}} +  q(x;t)\sqrt{\frac{e^{-\phi(x-dx)}}{e^{-\phi(x)}}} \right )}^{\rm leakage~from~state~{\it x}~to~its~neighbors} \nonumber \\ \label{eq:discrete}
\end{eqnarray}
\end{widetext}
The first term in Eq.~\ref{eq:discrete} corresponds to the probability flow into state $x$ from its neighbors $x-dx$ and $x+dx$ while the second terms corresponds to flow out of state $x$ into its neighbors $x-dx$ and $x+dx$.

In order to see the continuous-space limit of Eq.~\ref{eq:discrete}, assume that $q(x;t)$ and $\phi(x)$ are differentiable functions in $x$. Expanding the right hand side of Eq.~\ref{eq:discrete} as a Taylor series up to two orders in $dx$, we have
\begin{eqnarray}
\frac{\partial q(x;t)}{\partial t} &=& \mu dx^2 \left ( \frac{\partial^2 q(x;t)}{\partial x^2}  + \frac{\partial }{\partial x} \left [q(x;t) \cdot \frac{d \phi(x)}{d x}\right ]\right ) \nonumber \\
        &=& D\left ( \frac{\partial^2 q(x;t)}{\partial x^2}  - \frac{\partial }{\partial x} \left [q(x;t) \cdot \mathcal F(x) \right ]\right )  \label{eq:smol1}
\end{eqnarray}
where $D=\mu dx^2$ is the diffusion constant and we have identified $\mathcal F(x) = -\frac{d\phi(x)}{dx}$ as a {\it restoring force}.  The continuum-limit exists only when the rate $\mu$ scales such that $\mu dx^2$ is constant.  Eq.~\ref{eq:smol1} is valid not only for equilibrium situations ($\phi(x) = \beta G(x;\beta)$, where $G(x;\beta)$ is a free energy landscape and $\beta$ is inverse temperature) but also for non-equilibrium steady states (NESS).

\section*{Appendix III: MD simulation}
We performed a molecular dynamics simulation on $233$ water molecules~\citep{tip32,tip3mod} at 300K and at a constant volume using NAMD~\citep{phillips2005scalable} with help of the Langevin thermostat. The oxygen atom of one of the water molecules was fixed at the origin. The time step of integration was 1 fs and the trajectory was stored every 5 fs. Sampling the trajectory every 5 fs ensures that correlations in $n(t)$ haven't vanished.

%

\end{document}